%% file: scheffler.tex
\documentclass{appolb}
\usepackage{amsmath}
\usepackage{amsfonts}
\usepackage{epsfig}

\newcommand{\Z}{\ensuremath{\mathbb Z_3}}
\newcommand{\Zt}{\Z\ transformation}

\begin{document}
\title{PNJL model analysis of the Roberge-Weiss transition endpoint at imaginary chemical potential\thanks{Presented at the HIC for FAIR Workshop and XXVIII Max Born Symposium ``Three Days on Quarkyonic Island'', Wroc\l aw, Poland, May 19-21, 2011.}}
\author{David Scheffler, Michael Buballa, Jochen Wambach
\address{Institut f\"{u}r Kernphysik, Technische Universit\"at Darmstadt, Germany}
}

\maketitle

\begin{abstract}
Motivated by lattice QCD studies we investigate the RW transition endpoint at imaginary chemical potential in a two-flavor PNJL model. We focus on the quark-mass dependence of the endpoint using different forms of the Polyakov-loop potential.
\end{abstract}

  
\input{content.tex}


\end{document}

%% file: content.tex
\section{Introduction}
At imaginary chemical potential QCD has an interesting symmetry, known as the Roberge-Weiss (RW) symmetry~\cite{RW}: As a remnant of the \Z{} symmetry of the pure SU(3) gauge theory, certain shifts in the imaginary chemical potential $\mu = i \theta T$ can be undone by a \Zt{} leading to a periodicity of $\theta \rightarrow \theta+ 2\pi k/3$ with integer $k$ in thermodynamic quantities like the pressure.
At large temperatures the system undergoes a first-order transition jumping between different \Z{} sectors when crossing $\theta = (2 k+1) \,\pi/3$ for fixed temperature. Due to the periodicity $\theta$-even quantities show a cusp, whereas $\theta$-odd quantities have a jump. For low temperatures this transition is a crossover. In between there must be an endpoint of the RW transition which can be of first or second order. If the transition along $\theta = \pi/3$ ends in a first-order transition, there must be first-order lines departing from it implying that the endpoint is a triple point. As first-order phase transitions and second-order endpoints might influence the phase structure at real chemical potential, this warrants further studies.

Recent lattice QCD simulations at imaginary chemical potential for two and three quark flavors have shown that the order of the RW endpoint depends on the quark masses~\cite{DEliaRWEndpoint, PhilipsenRWEndpoint}. For low and high masses, the transition is of first order. A first-order transition at large quark masses is to be expected from the limit of pure SU(3) gauge theory. In the intermediate mass range the transition changes to second order with tricritical points in between.

Since lattice studies are hampered by the sign problem and are computationally very demanding, it is worth studying these aspects in effective models. In the Polyakov-loop extended Nambu--Jona-Lasinio (PNJL) model, which can be applied for real as well as at imaginary chemical potentials, we thus investigate the phase structure in the $\mu^2-T$-plane. At imaginary quark chemical potential the PNJL model also features the RW symmetry and we find the RW periodicity as well as the RW phase transition. 

The PNJL model at imaginary chemical potential has already been investigated by Sakai et.\,al.~\cite{Sakai0902,Sakai0904}. 
In a two-flavor PNJL model we extend their work and study the order of the RW phase transition endpoint for different Polyakov-loop potentials and analyze its dependence on the relative strength of the potentials~\cite{MScThesis}. This is done in two ways: Since quarks with larger mass have a smaller contribution to the pressure, increasing the quark mass makes the gluonic part more important. Alternatively, we directly change the prefactor of the gluonic contribution.

\section{Model}
We employ the PNJL model for two light quark flavors at real and imaginary chemical potential in mean-field approximation following the standard procedures. The Lagragian is given by
\begin{align*}
 \mathcal{L}_\text{PNJL} = &\bar\psi \left( i \gamma_\mu D^\mu - m_0 \right)\psi
+\frac{g_S}{2}\left[ (\bar\psi \psi)^2 + (\bar\psi i \gamma_5 \tau_a \psi)^2 \right]+ \mathcal{U}(\Phi,\bar\Phi)
\end{align*}
with quark fields $\psi$, covariant derivative $D^\mu$, bare quark mass $m_0$, coupling constant $g_S$ of the four-quark interaction and the Polyakov-loop potential $\mathcal{U}$, modelling the gluonic contributions which depends on the Polyakov-loop variables $\Phi$ and $\bar\Phi$. Parameters are taken from~\cite{Sakai0902}.

The \textit{extended \Zt}~\cite{Sakai0904} is given by
\begin{gather*}
\begin{split}
\theta&\quad\rightarrow\quad \theta+ 2\pi k/3 \\
\Phi &\quad\rightarrow\quad \Phi \exp{[-i 2\pi k/3]} \quad\text{with}\quad k \in \mathbb{Z}\text.
\end{split}
\label{eq:eZt}
\end{gather*}
A convenient definition is the \textit{modified Polyakov loop}, $\Psi = \Phi \exp{[i\theta]}$, which is then invariant under the extended \Zt. It can easily be shown that the PNJL model is invariant under the extended \Zt{} and thus possesses the RW periodicity at imaginary chemical potential.

\section{Results}
We start with a logarithmic form of the Polyakov-loop potential~\cite{RRW06},
\begin{equation*}
\frac{\mathcal{U}_\text{log}}{T^4} = -\frac{a(T)}{2} \Phi\bar\Phi + b(T) \log\left[ 1-6\Phi\bar\Phi + 4(\Phi^3+{\bar\Phi}^3) - 3(\Phi\bar\Phi)^2 \right] \text.
\end{equation*}

\begin{figure}
\centering
 \includegraphics[width=0.4\textwidth]{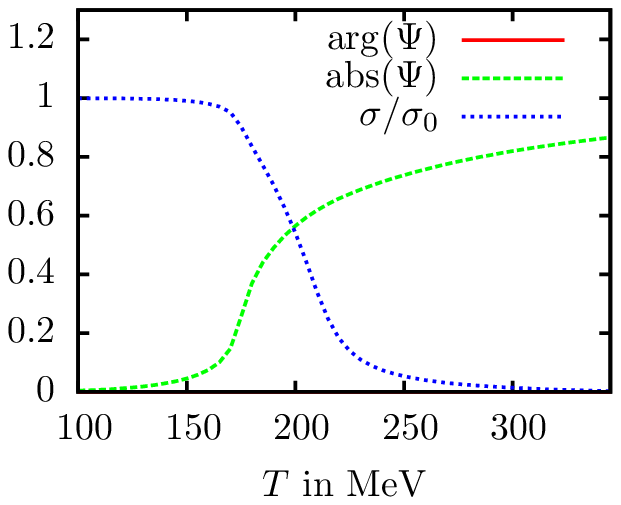}\hspace{1cm}
 \includegraphics[width=0.4\textwidth]{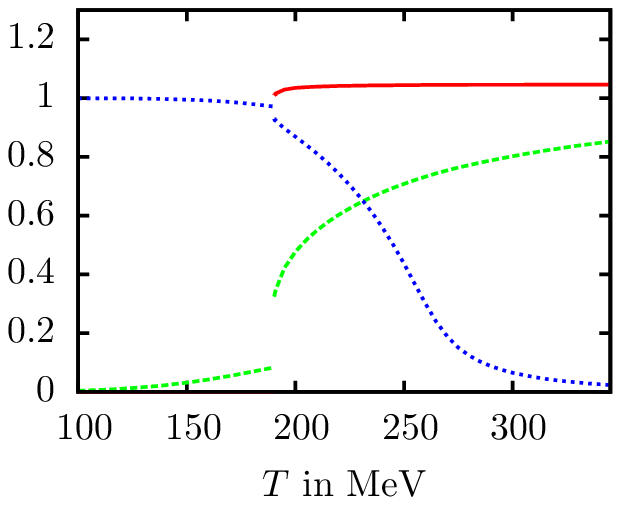}

\caption{Modified Polyakov-loop variables and the normalized chiral condensate $\sigma/\sigma_0$ at $\theta= 0$ (left) and $\theta= \pi/3$ (right) as functions of temperature. The phase of $\Psi$ vanishes at $\theta=0$ and only the positive branch is shown at $\theta=\pi/3$.}
\label{fig:orderparameters01}
\end{figure}

\begin{figure}
 \includegraphics[width=\textwidth]{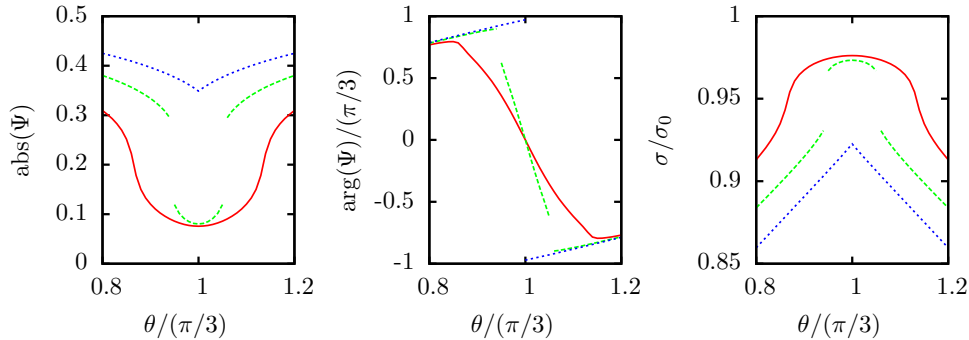}
 \caption{Dependence of the modified Polyakov-loop variables and the normalized chiral condensate on $\theta$ for different temperatures around $T_{RW}=190.3$ MeV (red solid: $T=185$~MeV, green dashed: $188$~MeV, blue dotted: $191$~MeV).}
\label{fig:orderparametersRW}
\end{figure}

\begin{figure}
\includegraphics[width=0.49\linewidth]{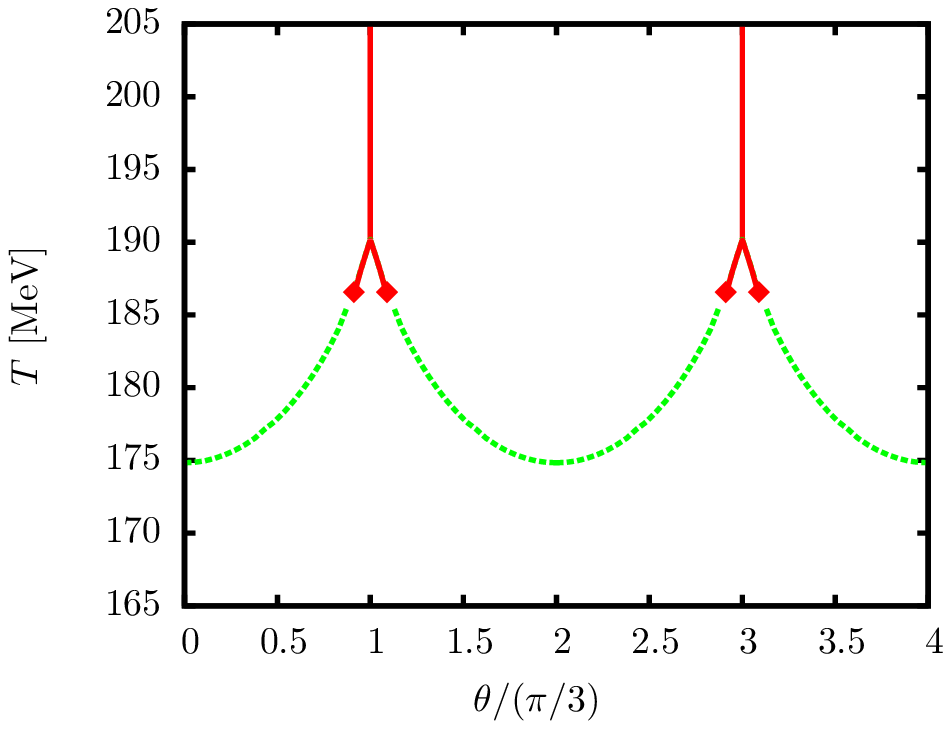}
\includegraphics[width=0.50\linewidth]{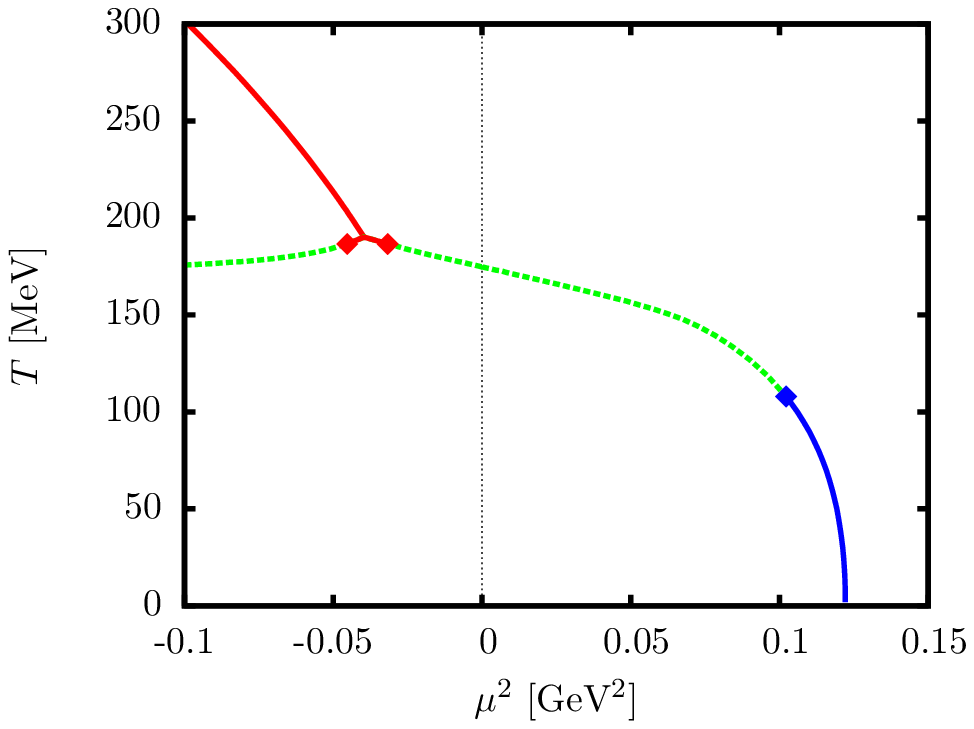}
\caption{Phase diagram in the $\theta-T$ (left) plane and the $\mu^2-T$ (right) plane. Red solid lines denote first-order RW/deconfinement transitions, green dashed lines show the deconfinement crossover, and the blue solid line at real chemical potential denotes the chiral first-order transition. The diamonds represent second-order endpoints.}
\label{fig:pd}
\end{figure}

We show the behavior of the order parameters at fixed $\theta=0$ and $\theta=\pi/3$ in Fig.~\ref{fig:orderparameters01}. Along $\theta=\pi/3$ which is in the middle of the period we find a jump in the absolute value and the phase of the Polyakov loop.
The dependence on $\theta$ at fixed temperatures close to the RW transition is displayed in Fig.~\ref{fig:orderparametersRW}. At temperatures higher than the transition temperature $T_{RW}$ the phase has a jump and the absolute value a cusp when crossing the RW phase transition, signalling the jump from one \Z{} sector to another. At temperatures slightly lower than $T_{RW}$ we however find two jumps in the phase and also in the absolute value. In addition the chiral condensate picks up the same discontinuities as the absolute value of the Polyakov loop. For even lower temperatures all transitions are continuous.
We summarize these findings in the PNJL phase diagram shown in Fig.~\ref{fig:pd}. Crossover lines are determined by the inflection point of the Polyakov-loop absolute value as a function of temperature. Chiral crossover lines are ommitted as they are not relevant for our current analysis.

\begin{figure}
 \includegraphics[height=0.26\textheight]{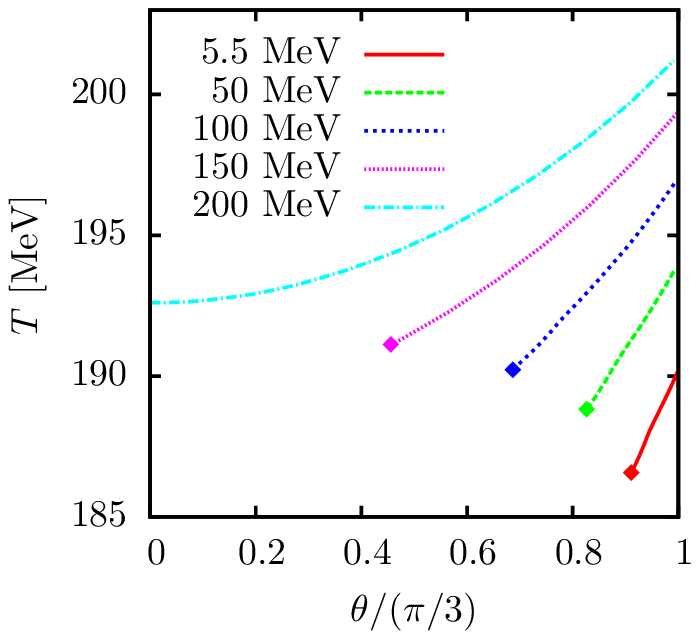}
 \includegraphics[height=0.26\textheight]{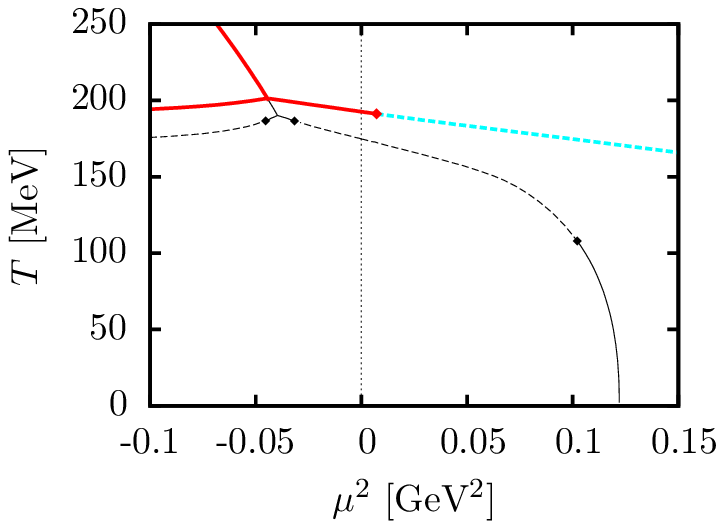}

\caption{Left panel: ``RW legs'' in the $\theta-T$ phase diagram for different values of the bare quark mass $m_0$. Right panel: Phase diagram in the $\mu^2-T$ plane for two different values of the bare quark mass. Red solid (blue dashed) lines show first-order RW/deconfinement (deconfinement crossover) transitions for a high bare quark mass $m_0 = 200$ MeV. Thin lines show the RW, deconfinement and chiral transitions for the standard value of $m_0 = 5.5$ MeV. }
\label{fig:legsvarM0}
\end{figure}

Using the logarithmic parametrization we find the RW endpoint to be a triple point independent of the quark masses, contrary to lattice results.  First-order lines departing from the triple point are nicely visible.
Increasing the bare quark masses $m_0$ leads to larger effective quark masses. This results in growing ``RW legs'', see Fig.~\ref{fig:legsvarM0}. For $m_0$ larger than about $180$~MeV the first-order lines even reach across the $\mu=0$~axis. This scenario is shown in the right panel of Fig.~\ref{fig:legsvarM0} in comparison to the standard-parameter results.

If instead the coupling constant $g_S$ is increased, which likewise leads to larger constituent quark masses, the same effect is found~\cite{Morita}.

Next we analyze the behavior of other Polyakov-loop potentials. Though all parametrizations are designed to reproduce pure-gauge lattice thermodynamics their effect on the RW endpoint is quite different.

The polynomial parametrization~\cite{RTW05} leads to a second-order transition for all examined quark masses. The reason is, that the polynomial parametrization shows a much weaker first-order transition in the heavy-quark limit.

\begin{figure}
\centering
 \includegraphics[width=0.55\textwidth]{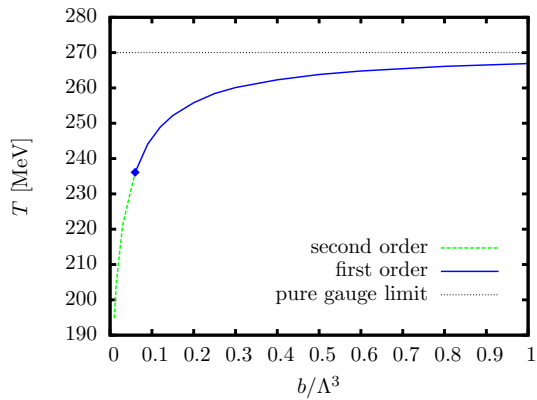}
 \caption{Temperature and order of the RW transition as function of parameter~$b$.}
 \label{fig:varb}
\end{figure}
Similarly, the Fukushima-type Polyakov-loop potential~\cite{Fukushima}, given by
\begin{equation*}
\frac{\mathcal{U}_\text{Fuku}}{T^4} = -b T \left( 54 e^{-a/T} \Phi\bar\Phi + \log\left[1 -6\Phi\bar\Phi + 4(\Phi^3+{\bar\Phi}^3) - 3(\Phi\bar\Phi)^2 \right] \right) \text,
\end{equation*}
produces a second-order transition for small quark masses and changes to first order only for very high quark masses where the PNJL model is not applicable any more. An alternative way to drive the system towards the pure gauge limit is to increase the global factor $b$ of the Fukushima-type Polyakov-loop potential. As presented in Fig.~\ref{fig:varb}, the RW endpoint changes from second to first order at about $b = 0.09 \Lambda^3$ whereas the default value for $N_f=2$ is $b=0.015 \Lambda^3$. For $b>0.5 \Lambda^3$ the ``RW legs'' reach across the temperature axis. With increasing $b$ the transition temperature approaches the heavy-quark limit of $T_c=270$~MeV.

We conclude, that the PNJL model together with currently available parametrizations for the Polyakov-loop potential is not able to reproduce the mass dependence found in lattice QCD studies. Sakai et.\,al.\ have shown that the 'entanglement' PNJL (EPNJL) model, which uses a Polyakov-loop dependent coupling $g_S$, reproduces the desired behavior~\cite{SakaiEPNJL}.

\section{Summary}
We have shown that the choice of the Polyakov-loop potential pa\-ram\-e\-tri\-za\-tion has an important influence on the order of the RW phase transition endpoint. Modifying the strength of the quark degrees of freedom relative to the Polyakov-loop potential which models the gluon degrees of freedom, we find interesting changes in the phase structure at imaginary and real chemical potential. Results from lattice QCD should be used to constrain the Polyakov-loop potential parametrizations used in model studies.

\bigskip
The authors thank the organizers for an interesting workshop. D.\,S.\ acknowledges travel support by HIC for FAIR. This work was partially supported by the German Federal Ministry of Education and Research under project nr.\ 06DA9047I, the Helmholtz Alliance EMMI and the Helmholtz International Center for FAIR.